\documentclass[twocolumn]{IEEEtran}
\usepackage[noadjust]{cite}
\usepackage[cmex10]{amsmath}
\usepackage{amsfonts}
\usepackage{amssymb}
\usepackage{algorithm}
\usepackage{algorithmic}
\usepackage[dvipsnames]{xcolor}
\usepackage[colorlinks,linkcolor=black,anchorcolor=black,citecolor=black]{hyperref}
\usepackage{graphicx}
\usepackage{verbatim}
\usepackage{indentfirst}
\usepackage{epsfig}
\usepackage{bm}
\usepackage{multirow}
\usepackage{subfigure}
\usepackage{setspace}
\usepackage{makecell}
\usepackage{svg}

\IEEEoverridecommandlockouts

\begin{document}
\title{Integrating Base Station with Intelligent Surface for 6G Wireless Networks: Architectures, Design Issues, and Future Directions}
\author{Yuwei Huang, {\it Student Member, IEEE}, Lipeng Zhu, {\it Member, IEEE}, and Rui Zhang, {\it Fellow, IEEE}

\thanks{

Y. Huang and L. Zhu (corresponding author) are with National University of Singapore; R. Zhang (corresponding author) is with The Chinese University of Hong Kong, China, Shenzhen Research Institute of Big Data, China, and National University of Singapore, Singapore.
}
}
\maketitle

\begin{abstract}
Intelligent surface (IS) is envisioned as a promising technology for the sixth-generation (6G) wireless networks, which can effectively reconfigure the wireless propagation environment via dynamically controllable signal reflection/transmission. In particular, integrating passive IS into the base station (BS) is a novel solution to enhance the wireless network throughput and coverage both cost-effectively and energy-efficiently. In this article, we provide an overview of IS-integrated BSs for wireless networks.  Specifically, we present three different practical architectures based on the integrated location of IS and compare them from several key design aspects. Then, the main design issues are discussed for all architectures, which include channel modelling, channel state information (CSI) acquisition, and IS passive reflection/transmission design. Moreover, numerical results are presented to compare the performance of different IS-integrated BS architectures as well as the conventional BS without IS. Finally, promising directions are pointed out to stimulate future research on IS-BS/terminal integration in wireless networks.
\end{abstract}

\section{Introduction}
To support emerging applications like immersive communication, extended reality, and tactile Internet, sixth-generation (6G) wireless networks are expected to significantly exceed the capabilities of today's fifth-generation (5G) networks \cite{6G}. In particular, future wireless communication systems should deliver services with extremely high throughput, ultra-high reliability, minimal latency, and extremely low power consumption. However, existing 5G technologies such as massive multiple-input-multiple-output (MIMO) and millimeter-wave (mmWave) communications may not be able to meet these demands efficiently, since they enhance the system performance at the cost of increasingly higher energy consumption and hardware cost. In addition, the performance of existing wireless networks is fundamentally limited by various impairments in wireless channels, such as path loss, shadowing, and multi-path fading, which are influenced by random environmental factors. To tackle these issues, the promising technology of {\it Intelligent Surface} (IS) has emerged, which is a software-controlled metasurface equipped with massive reflective/transmissive elements. Notice that these reflective/transmissive elements can individually tune the phase shift and/or amplitude of the incident signal in real-time, so as to reconfigure the radio propagation environment for enhancing the wireless network performance \cite{tutorial}. As IS typically operates in passive signal reflection/transmission, it dispenses with radio frequency (RF) chains, which thus leads to significantly lower hardware cost and energy consumption as compared to traditional active antenna arrays. Owing to the benefits introduced by the IS, it has been applied to achieve various functions in wireless networks, such as passive beamforming, interference nulling/cancellation, channel distribution refining, and so on. In general, the IS can represent the well-known intelligent reflecting/refracting surface (IRS) and reconfigurable intelligent surface (RIS) as well as other metamaterial-based technologies \cite{tutorial}.

\begin{figure*}
\centering
\includegraphics[width=15cm]{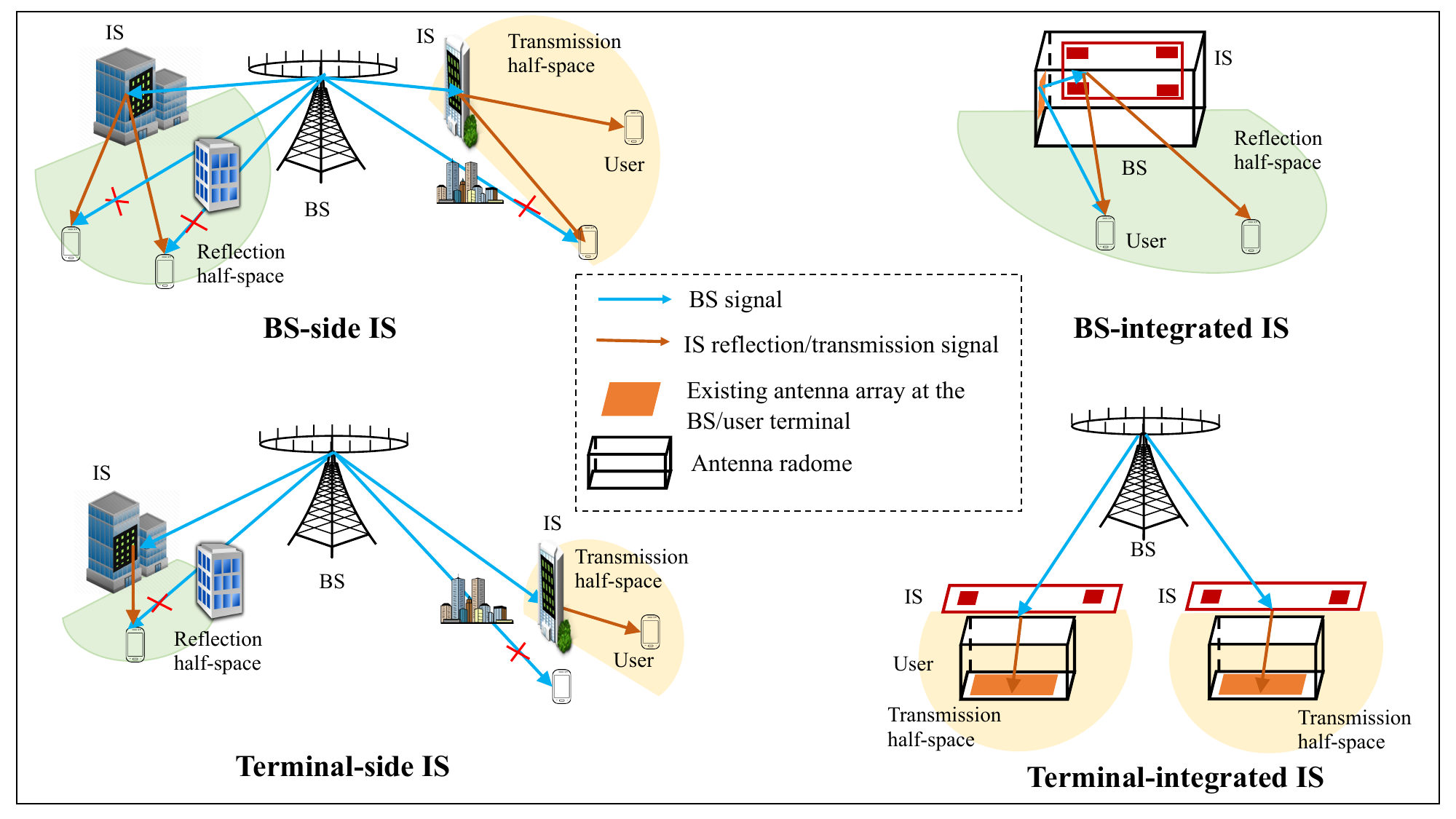}
\caption{Illustration of different IS deployment strategies, i.e., BS/terminal-side IS deployment and BS/terminal-integrated IS deployment.}\label{deployment}
\end{figure*}

To effectively achieve the high-performance gains of IS, it is crucial to strategically deploy the IS between the base station (BS) and user terminals to ensure the avoidance of high product-distance path loss of the cascaded BS-IS-user channel \cite{deploy}, which corresponds to the {\it product} of the path losses in the user-IS and IS-BS channels. These path losses are proportional to their respective signal propagation distances. Depending on its deployed location, the IS's deployment strategies can be generally classified into the following four categories: BS-side IS, BS-integrated IS, terminal-side IS, and terminal-integrated IS. As shown in Fig. \ref{deployment}, for {\it BS/terminal-side IS}, the IS is deployed in the environment separately from the BS/user terminal, but as close to the BS/user terminal as possible (e.g., in the order of hundreds to thousands of carrier wavelengths) to reduce the product of its distances with both of them \cite{co_site, user_side}. To further reduce the distance between the IS and BS/user terminal (e.g., to the order of several to tens of carrier wavelengths), the {\it BS/terminal-integrated IS} has been proposed to integrate the IS along with the active antennas of the BS/user terminal within the same antenna radome \cite{integrated_IS,integrated_IS_codebook}. Furthermore, for each of the above deployment strategies, the IS in general can operate in either the {\it reflective} or {\it transmissive} mode, for the BS and user terminals located at the same side or different sides of the IS, respectively (see Fig. \ref{deployment}), so that the signal can be reflected by or pass through it.

Generally speaking, deploying the IS closer to the BS can reduce the IS deployment cost and improve the system performance more effectively as compared to its counterpart closer to user terminals, even under the same product-distance of the cascaded user-IS-BS channel. This is because the BS is usually at a fixed location, and thus the IS deployed near the BS can potentially serve more user terminals randomly distributed in its reflection/transmission half-space (see Fig. \ref{deployment}). Moreover, BS-integrated IS can further reduce the distance between the BS's antenna array and IS, thereby minimizing the product-distance path loss in the cascaded channel and facilitating easier real-time control of the IS by the BS. Furthermore, as compared to the conventional multi-antenna BS without IS, the BS-integrated IS can enhance the BS's communication throughput and coverage performance without the need to deploy more active antennas, thus providing a practically appealing low-cost BS architecture for future wireless networks such as 6G.

Despite these advantages, the BS-integrated IS introduces new design issues that must be addressed to fully realize its potential for improving wireless network performance. For example, the extremely short distance between the BS's antenna array and IS causes their complicated signal interactions (i.e., near-field channel condition as well as the intricately coupled single-reflection/transmission and double-reflection/transmission signals), which need to be well characterized to model the resulting cascaded channels accurately. Moreover, the acquisition of channel state information (CSI) between the BS and user terminals with integrated ISs, which is essential for reaping the IS's maximal performance gains, becomes more challenging, thus requiring new approaches to solve this issue efficiently, for both reflection/transmission-based ISs.

Motivated by the above, this article aims to provide a comprehensive overview of IS-integrated BS architectures for 6G wireless networks. The rest of this article is organized as follows. Section \ref{architecture} introduces and compares three different IS-integrated BS architectures based on the integrated location of IS. Section \ref{design} discusses the main design issues for IS-integrated BS architectures, including channel modelling, CSI acquisition, and IS passive reflection/transmission design. Section \ref{simulation} presents numerical results to evaluate the performance of our considered IS-integrated BS architectures with different IS passive reflection/transmission designs, as well as with the conventional multi-antenna BS without IS. Finally, in Section \ref{future}, promising directions are pointed out for the IS-BS/terminal integration in wireless networks to motivate future research.

\section{IS-Integrated BS Architectures}\label{architecture}
In this section, according to the integrated location of IS, we classify the IS-integrated BS architectures into three different categories: backside-IS architecture, frontside-IS architecture, and surrounding-IS architecture. In the following, we elaborate the key design aspects of each architecture, including the number of surfaces, IS working mode, number of signal hops, surface scalability, power loss, and performance gain. Notice that the surface scalability typically refers to the ability to scale or modify the surface area of the antenna radome to deploy IS elements. When the surface scalability is good, the number of deployable IS elements is large. By contrast, when the surface scalability is limited, the number of deployable IS elements is small.
\begin{figure*}
\centering
\includegraphics[width=15cm]{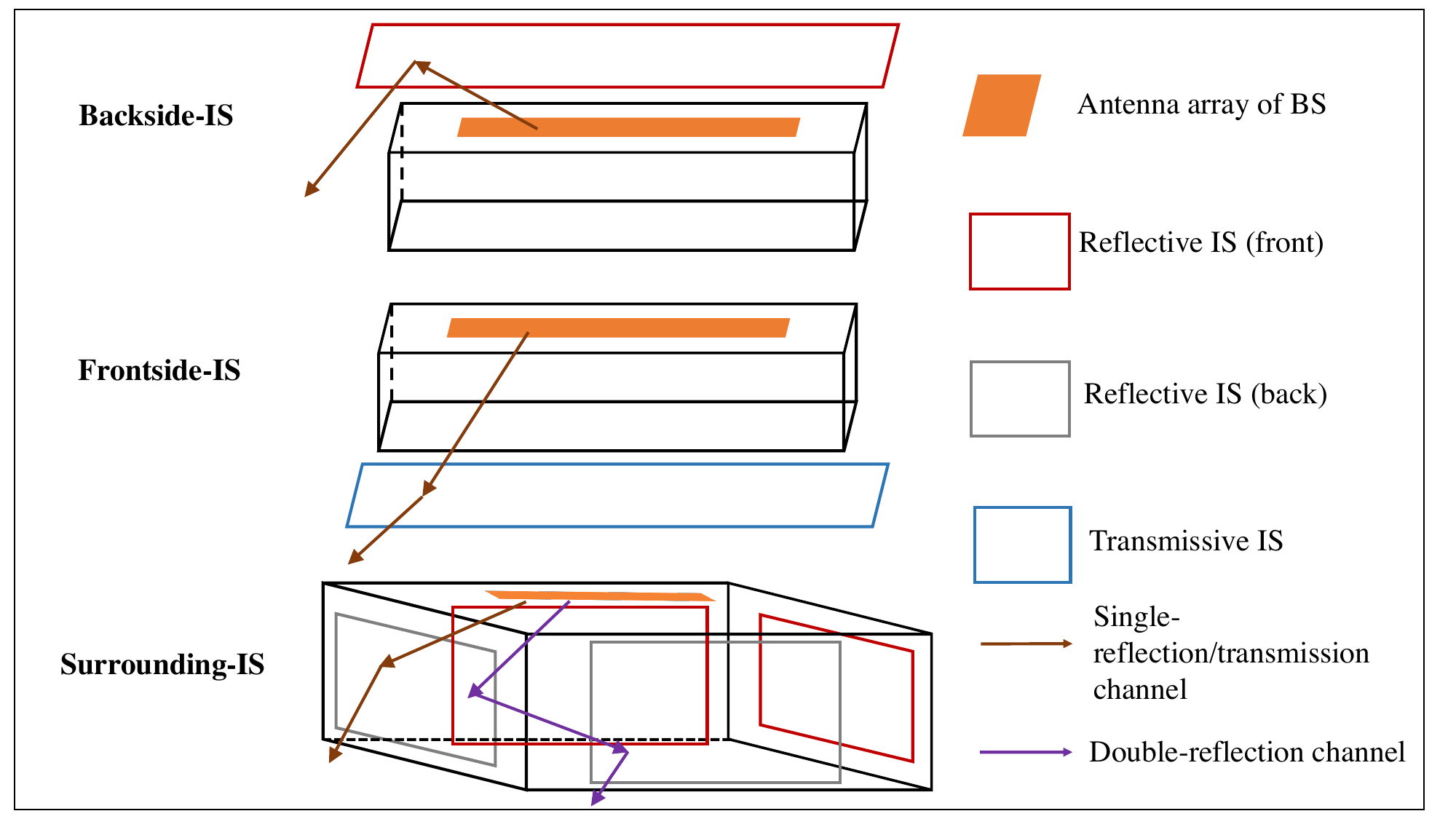}
\caption{Illustration of different IS-integrated BS architectures, i.e., backside-IS, frontside-IS, and surrounding-IS architectures.}\label{integrated_architecture}
\vspace{-10pt}\end{figure*} 

As depicted in Fig. \ref{integrated_architecture}, the {\it backside-IS architecture} integrates a single planar surface mounted with {\it reflective} elements at the backside of the BS's antenna array, which bears similarity to the traditional reflectarray antennas \cite{reflectarray}. Since the IS is usually deployed outside the antenna radome of the BS under this architecture, the number of deployable IS elements can be increased without being limited by the size of BS's antenna radome, which, however, increases the propagation loss due to longer signal propagation distances between the IS elements and BS antennas. Besides, only single-reflection channels exist under this architecture, which may constrain the IS reflection gain as compared to other architectures with multiple signal reflections/transmissions under the same total IS surface aperture.

The second category is the {\it frontside-IS architecture}, which integrates the {\it transmissive} IS in front of the BS's antenna array, as depicted in Fig. \ref{integrated_architecture}. Notice that when the transmissive coefficients of all IS elements are fixed, this architecture is similar to traditional electromagnetic (EM) lens-based antennas \cite{lens}. Analogous to the backside-IS architecture, the IS size for the frontside-IS architecture is not limited by the size of BS's antenna radome since the IS can be deployed outside the BS's antenna radome. Nevertheless, the signal penetration loss under this architecture is usually higher than the signal reflection loss of the backside-IS architecture, which constrains the IS passive beamforming gain. Note that the existing literature \cite{transmissive_irs} has proposed to integrate multi-layer {\it stacked} transmissive ISs in front of the BS's antenna array to  aid in its communication with user terminals. In this case, as the number of layers increases, the IS passive beamforming gain is boosted owing to more IS elements deployed. However, this also increases the signal penetration loss, which degrades the system performance. Thus, there is an inherent trade-off in choosing the optimal number of layers for the multi-layer transmissive IS to maximize system performance. In this article, we focus on the single-layer transmissive IS for the frontside-IS architecture for simplicity.

Finally, {\it the surrounding-IS architecture} has been recently proposed in \cite{integrated_IS,integrated_IS_codebook} to integrate four {\it distributed} ISs and an antenna array within the antenna radome at the BS, as depicted in Fig. \ref{integrated_architecture}, where the ISs work on the {\it reflective} mode. Specifically, the BS's antenna array is positioned at the center of the top surface of the antenna radome, while the ISs are deployed on the left, right, front, and back surfaces, perpendicular to the antenna array. In this architecture, due to the extremely short distances between various ISs, the double-reflection channels involving two different ISs become crucial. These channels can be effectively combined with single-reflection channels to enhance passive beamforming gain of the IS and thus improve user rate performance. However, since the ISs are deployed at the side surfaces of the BS's antenna radome, the number of deployable IS elements is limited by the size of the BS's antenna radome in practice. 

To summarize, Table \ref{compare} compares the three IS-integrated BS architectures above in various aspects. In practice, the optimal IS-integrated BS architecture may be varied across different scenarios, which depends on many factors, such as operation environments, system requirements, and hardware constraints. Thus, more research efforts are needed to develop customized architectures and algorithms for IS-integrated BSs. In this article, we will provide numerical results to compare the performance for all IS-integrated BS architectures under the same total surface aperture of the IS, as shown in Section \ref{simulation}.

\begin{table*}\scriptsize
\centering
\caption{Comparison of three IS-integrated BS architectures}\label{compare}
\begin{tabular}{|c|c|c|c|c|c|c|}
\hline
Architecture&Number of surfaces&IS working mode&Number of signal hops&Surface scalability& Power loss&\makecell[c]{Performance gain\\ (with the same total surface aperture)}\\
\hline
Backside-IS&Single&Reflective&Single&Good&Moderate&Moderate\\
\hline
Frontside-IS&Single/Multiple(stacked)&Transmissive&Single/Multiple&Good&Moderate/High&High\\
\hline
Surrounding -IS&Single/Multiple(distributed)&Reflective&Single and double&Limited&Low&Very High\\
\hline
\end{tabular}
\end{table*}

\section{Design Issues for IS-Integrated BS}\label{design}
In this section, we outline the intricate design issues in IS-integrated BS-assisted wireless communications. Additionally, we explore promising solutions aimed at addressing these challenges to stimulate future research efforts.

\subsection{Channel Modelling}
\begin{figure}
\centering
\includegraphics[width=9cm]{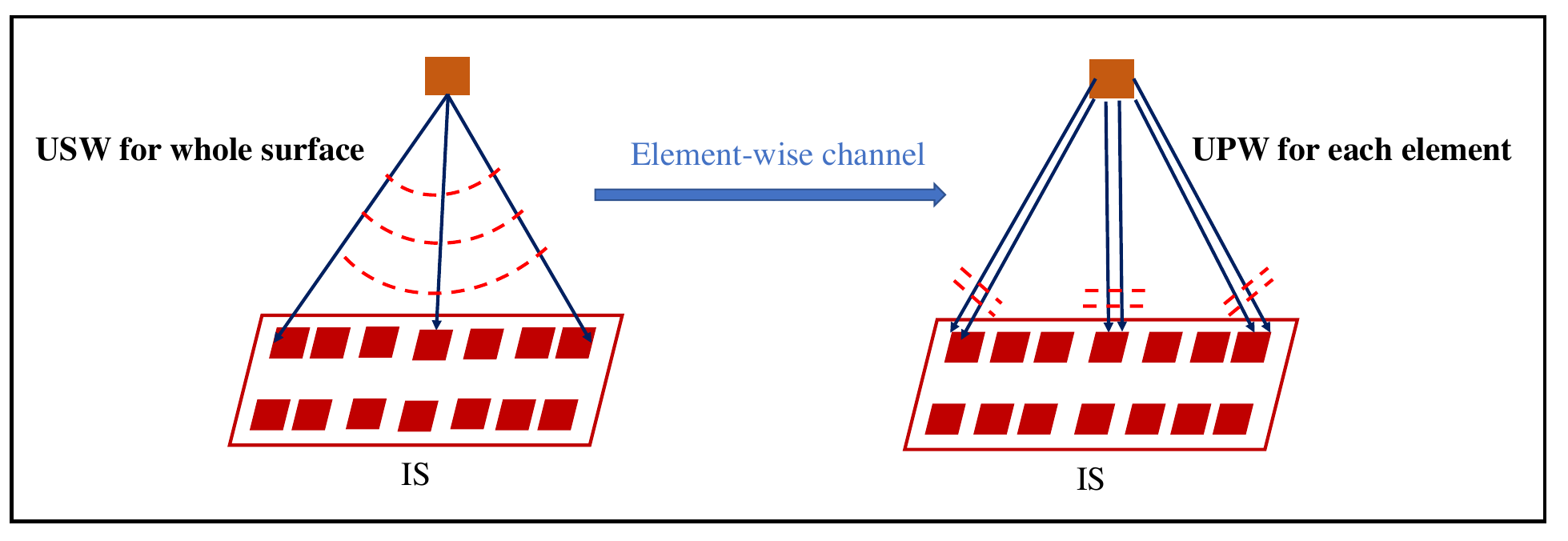}
\caption{Illustration of element-wise channel model.}\label{element}
\vspace{-10pt}\end{figure} 
For the user-BS and user-IS channels in the IS-integrated BS architectures, since the signal propagation distance between the BS/IS and user terminal is much longer compared to the size of BS/IS, the channel condition falls into the far-field region. Therefore, uniform plane wave (UPW) propagation can be used for channel modelling \cite{channel_model}. In this case, all antennas of the BS or elements of the IS share the same amplitude and angle-of-arrival/departure (AoA/AoD) for the same channel path from the user terminal. However, due to the close deployment locations of the integrated ISs, the signal propagation distance between the BS's antenna array and IS, and that between different ISs for the surrounding-IS architecture, are much shorter compared to the size of IS, which renders that the UPW assumption cannot hold for these involved channels. To tackle this issue, the uniform spherical wave (USW) propagation model is adopted for the IS-BS and inter-IS channels to take the variation of signal amplitude/phase across all BS's antennas/IS's elements into consideration \cite{channel_model}. Fortunately, the far-field channel condition still holds between each BS antenna and each IS element, and the elements in different ISs for the surrounding-IS architecture \cite{integrated_IS}, where the UPW model can be applied to describe the element-wise channel between each IS element and each antenna of the BS, as well as between elements in different ISs under the surrounding-IS architecture, as depicted in Fig. \ref{element}. Specifically, each single-reflection/transmission channel path from a user terminal to an antenna of the BS via an element of the integrated IS can be modelled as the product of the channel between the IS's element and user terminal, the reflection/transmission gain provided by the IS element, the channel between the IS's element and BS's antenna, and the antenna gain provided by the BS. In addition, the double-reflection channel via two different ISs for the surrounding-IS architecture can be modelled in a similar way. Interested readers can refer to \cite{integrated_IS} for more details about the channel modelling of IS-integrated BS.

In practice, the reflection/transmission gain of each IS element is non-isotropic for different AoAs and AoDs, and it is also dependent on IS's aperture size  \cite{integrated_IS}. Besides,  note that the antenna radiation pattern of conventional BS without IS usually directs its main-lobe with high antenna gain forwards to serve user terminals, which thus results in low or negligible side-lobe antenna gain backwards. Thus, to enable the integrated ISs for enhancing the system performance, the BS's antenna radiation pattern should be carefully designed in the IS-integrated BS, which is dependent on the integrated location of the IS and the distribution of user terminals. For example, for the backside-IS architecture, the BS with much higher antenna gain backwards can enable the IS elements deployed at the backside of the BS's antenna array, so as to enlarge their reflection gain.

\subsection{CSI Acquisition}
Notice that maximizing the passive beamforming gain of IS requires accurate CSI acquisition. However, this presents practical challenges in IS-integrated BS architectures since the IS elements do not have active RF chains, particularly in the surrounding-IS architecture with inter-IS channels. The existing literature describes two  primary methods for IS channel estimation based on various IS configurations: {\it semi-passive IS} and  {\it fully passive IS}. In the first case involving semi-passive IS, sensing devices (such as low-power sensors) equipped with inexpensive {\it receive} RF chains are integrated into IS for the frontside-IS and backside-IS architectures, which allows separate estimation of BS-IS and user-IS channels by exploiting the pilot signals transmitted by the BS and/or user terminals \cite{co_site}. Nevertheless, due to the existence of additional inter-IS reflections under the surrounding-IS architecture, the CSI acquisition for both single- and double- reflection channels becomes more challenging, even with the help of sensing devices mounted on each IS.

On the other hand, for the case involving fully-passive IS, the cascaded BS-IS-user or user-IS-BS channels can be estimated at the user terminal/BS by exploiting the downlink/uplink pilots signals for the frontside-IS and backside-IS architectures, which, however, may incur high channel estimation overhead due to a large number of unknown channel parameters. For reducing the channel estimation overhead, some efficient strategies have been proposed in the existing literature (see \cite{channel_estimation_survey} and the references therein) for these two architectures, such as  IS element grouping, reference user-based channel estimation, anchor-aided channel estimation, channel estimation based on channel sparsity, and so on. 

While for the surrounding-IS architecture, although the single-reflection channels can be estimated by leveraging the channel estimation strategies for the frontside-/backside-IS architecture via sequentially turning on the ISs, the double-reflection channels over different pairs of ISs are more difficult to estimate since they are coupled and hard to estimate separately. Fortunately, thanks to the extremely short distances between BS's antenna array and ISs, and the fixed position/orientation of each IS in the surrounding-IS architecture, the single-/double-reflection channels can be expressed as functions of the AoAs at the BS and the complex coefficients of incident paths \cite{integrated_IS_codebook}. These function expressions can be derived offline using method such as element-wise channel modelling, the ray-tracing technique, or electromagnetic (EM) simulation. Based on the above, the AoAs and complex coefficients of all channel paths from each user terminal can be estimated in real-time at the BS by turning off all ISs. These estimates are then used to reconstruct the effective channel of each user terminal when the ISs are turned on. Notice that this channel estimation strategy is also applicable to the frontside-IS and backside-IS architectures with lower complexity to implement. Nevertheless, the acquisition of complex coefficients and AoAs of all incident channel paths is  difficult to implement compared to the conventional CSI acquisition for the composite channels only at the BS, and the cascaded single-reflection/transmission channel estimation for the backside-IS and frontside-IS architectures may still incur large channel estimation overhead when the number of IS elements and/or the number of user terminals becomes large. As such, more efforts are required for further investigation of efficient channel estimation strategies for the IS-integrated BS architectures.

\subsection{IS Passive Reflection/Transmission Design}
Depending on whether the CSI for all IS-associated channels is required, the IS passive reflection/transmission design can be divided into CSI-based and codebook-based approaches. In the CSI-based design, it is assumed that the CSI for all relevant channels is available at the BS. This is achieved through efficient channel estimation strategies tailored for IS-integrated BS architectures, as discussed in the previous subsection \cite{channel_estimation_survey}. In this case, the reflective/transmissive coefficients of all IS elements can be optimized jointly with the BS's active beamforming for enhancing system performance, such as maximizing the achievable sum-rate of all users, maximizing the minimum achievable rate among all users, and so on. Several algorithms have been proposed in the literature to address the formulated optimization problem \cite{user_side,co_site,integrated_IS}. For instance, the alternating optimization (AO) method, explored in \cite{user_side,co_site}, iteratively optimizes the active beamforming of the BS and passive reflective/transmissive coefficients of IS until convergence is reached. In addition, the successive refinement method, as applied in \cite{integrated_IS}, alternates updates of the reflective/transmissive coefficients for each IS element while keeping those of other IS elements fixed.

In contrast, the codebook-based IS passive reflection/transmission design does not require the explicit CSI for all IS-involved channels. Specifically, a codebook containing a collection of codewords (i.e., IS passive reflection/transmission patterns) is created offline and stored at the BS. Subsequently, for each codeword (i.e., IS passive reflection/transmission pattern), the BS estimates the effective channels with user terminals using their transmitted pilot signals and assesses the communication performance (such as the achievable sum-rate of all users). Finally, the codeword that yields the best communication performance is chosen for data transmission. The existing literature has proposed several methods for IS passive reflection/transmission codebook design \cite{integrated_IS,integrated_IS_codebook}. For example, a two-dimensional discrete Fourier transform (DFT) codebook has been adopted in \cite{integrated_IS} to search the optimal codeword, which, however, achieves worse performance for the IS-integrated BS architectures due to the near-field effects as well as the coupling between single- and double-reflection channels under the surrounding-IS architecture. Besides, the authors in \cite{integrated_IS} utilized a random codebook to randomly generate a given number of IS patterns, which, however, becomes inefficient for the surrounding-IS architecture due to the coupled terms in double-reflection channels \cite{integrated_IS}. To solve this issue, an iterative random phase algorithm (IRPA) for the surrounding-IS architecture was further proposed in \cite{integrated_IS}, where the reflective coefficients of multiple ISs are optimized based on the generated random codebook in an iterative manner. Since the random codebook-based and IRPA-based designs both need sufficiently high training overhead to approach the theoretical performance bound under the perfect CSI, they may reduce the effective data transmission time and thus compromise the achievable rate performance in practice. To address this challenge, a robust codebook design was proposed in \cite{integrated_IS_codebook} aimed at enhancing performance across the entire coverage area of the BS. In particular, the BS's coverage area is segmented into multiple non-overlapping sectors based on the azimuth angle, with a unique codeword (i.e., IS passive reflection/transmission pattern) designed for each sector to optimize its coverage performance. This efficient sector division strategy minimizes the number of codewords required in the robust codebook and allows the IS patterns to adapt slowly to wireless channels, thereby reducing training overhead significantly. 
\begin{figure*}
\centering
\includegraphics[width=15cm]{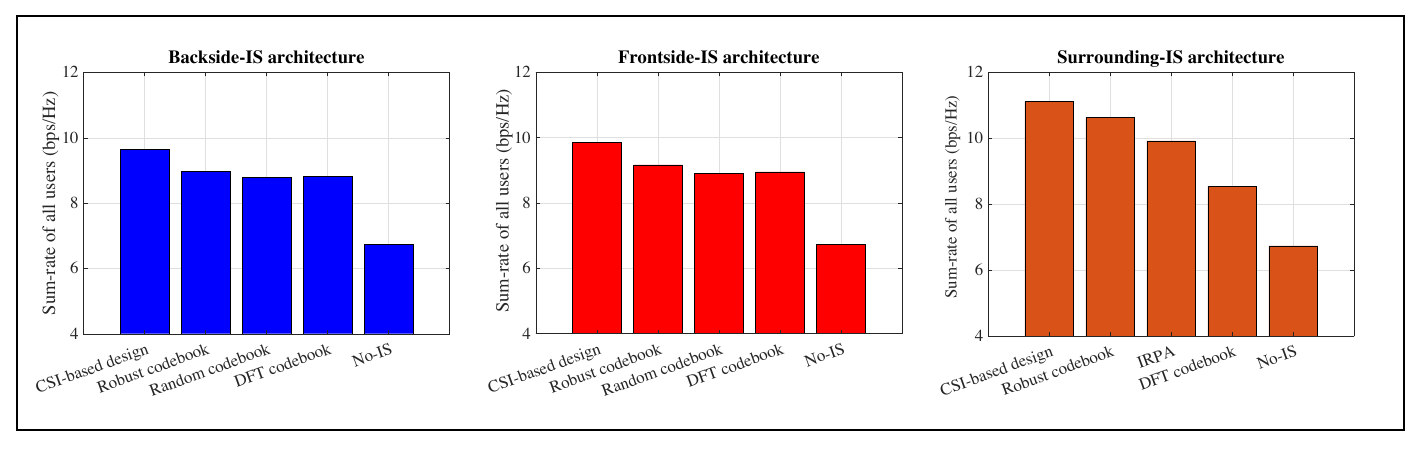}
\caption{Performance comparison for IS-integrated BS architectures with different IS passive reflection/transmission designs.}\label{results}
\end{figure*} 

\section{Numerical Results}\label{simulation}
Numerical results are presented in this section to compare the performances of all the three considered IS-integrated BS architectures under both CSI-based and codebook-based  (including DFT codebook, random codebook/IRPA, and robust codebook) IS passive reflection/transmission designs, as well as the conventional BS without IS (termed as ``no-IS'' scheme). In particular, we consider the uplink communication from four ground users to a four-antenna BS via the integrated ISs with 40 elements in total, where the BS's antenna radome is deployed at the altitude of $5~\text{meter (m)}$, the BS's antenna array faces the ground directly, and the user terminals are randomly located on the ground. The user-BS channels and user-IS channels are both assumed to follow Rician fading with Rician factor being $5~\text{dB}$. We assume that the BS and IS are both equipped with the uniform planar array (UPA). Besides, the carrier frequency is set as $6~\text{GHz}$, the antenna/IS element spacing is set as half-wavelength, and the area aperture of each IS element is set to be square of half-wavelength. In addition, the antenna radiation pattern of the BS is assumed to be half-isotropic, where the main-lobe is set backwards for the backside-IS architecture, and forwards for the other two architectures, while the IS reflection/transmission gain is modelled by adopting the model proposed in \cite{integrated_IS}. For the case of random codebook, the random codebook-based design is adopted for the backside-IS and frontside-IS architectures, while the IRPA-based design is adopted for the surrounding-IS architecture, where the total number of randomly generated codewords (i.e., IS passive reflection/transmission patterns) is set as 5000. Moreover, for the robust codebook-based design, the codeword for each sector is optimized to maximize the average effective channel power of all locations within it, and the overall codebook is formed by combining all IS codebooks designed for different number of sectors \cite{integrated_IS_codebook}, where the codebook size is set as 15. Notice that the training overhead for all codebook-based designs is the product of the number of user terminals and codebook size (i.e., 5000 for random/IRPA codebook and 15 for robust codebook). All results in the simulation are averaged over 100 random realizations of wireless channels and user locations.

Fig. \ref{results} shows the achievable sum-rate of all users for the three considered IS-integrated BS architectures under different strategies for IS passive reflection/transmission design. Notice that the successive refinement method is adopted for IS passive reflection/transmission design in CSI-based design, where the computational complexity is given in \cite{integrated_IS}. It is observed that  the IS-integrated BS architectures outperform the no-IS scheme. This is because the integrated ISs can provide additional reflection/transmission gains to enhance the system performance. Besides, under the same total surface aperture of IS, it is observed that the surrounding-IS architecture achieves higher rate performance than backside-IS and frontside-IS architectures. This is due to the fact that the additional double-reflection channels are exploited to enhance IS passive beamforming gains and reduce power loss (including both the propagation and reflection/penetration loss) under the surrounding-IS architecture. Furthermore, for each architecture, it is observed that the robust codebook-based design achieves higher user rate performance than DFT codebook-based and random codebook/IRPA-based designs. The reason lies in that the codewords in the robust codebook are designed to maintain performance consistency across all locations within the coverage area. Moreover, the robust codebook has a smaller codebook size and supports more stability of IS patterns due to its efficient sector-based design, thereby remarkably reducing real-time training overhead.

\section{Future Directions}\label{future}
\begin{figure*}
\centering
\includegraphics[width=15cm]{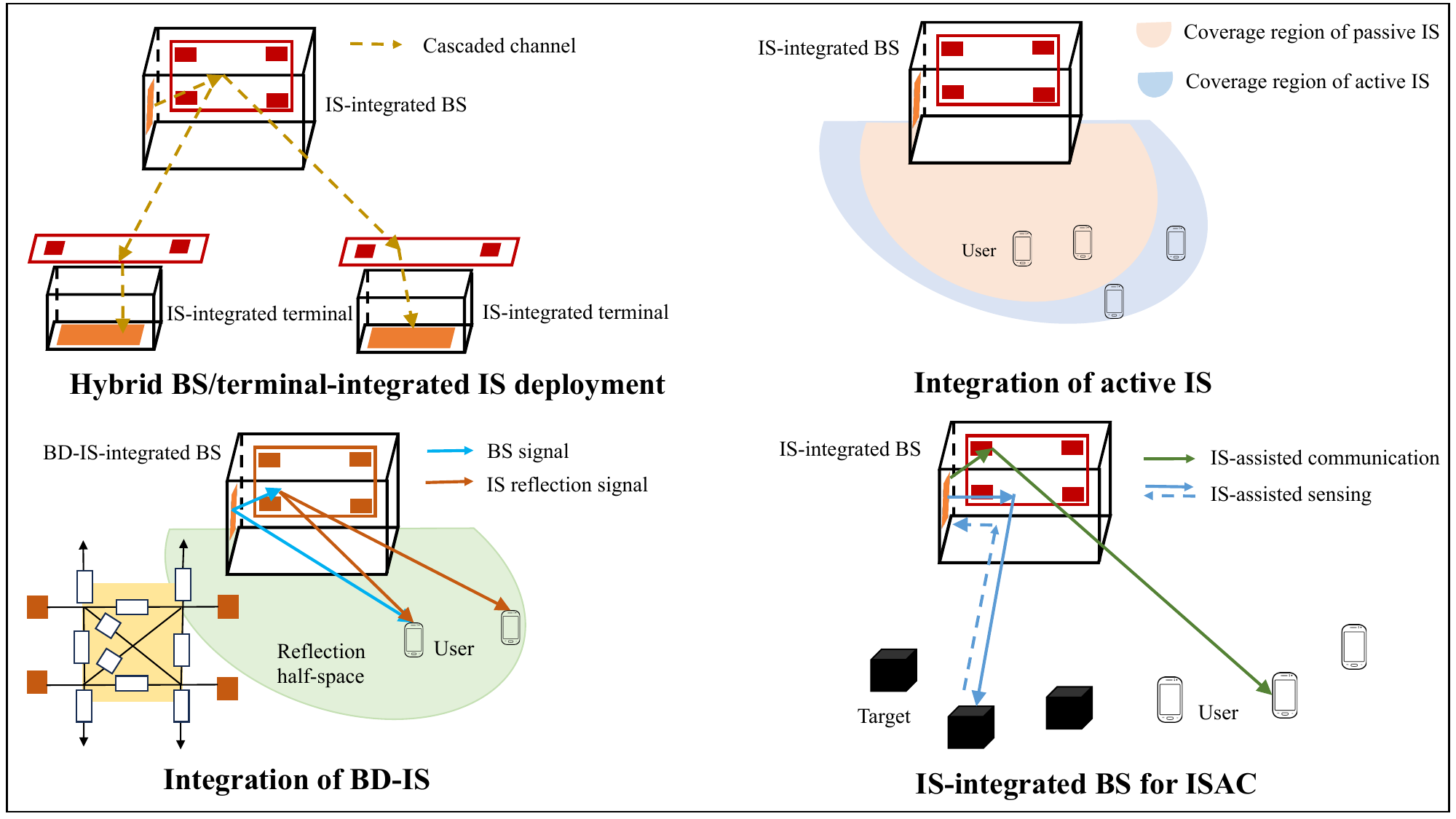}
\caption{Promising research directions for IS-BS/terminal integration in wireless networks.}\label{future_picture}
\end{figure*} 
In addition to the aforementioned design issues for IS-integrated BS in wireless communications, there are other important issues that merit further investigation in future work. Below, we highlight several promising directions to stimulate upcoming research endeavours.

\subsection{IS-Integrated Terminal  and Hybrid BS/Terminal-Integrated IS Deployment}

In practice, the presented three types of IS-integrated BS architectures can be extended to their user terminal counterparts. However, due to the much smaller size of the user terminal's antenna radome, the number of deployable IS elements for the IS-integrated terminal is limited. To acquire higher IS reflection/transmission gain, we can consider a hybrid BS/terminal-integrated IS deployment, as depicted in Fig. \ref{future_picture}, where the link between the IS-integrated BS and IS-integrated terminal can be exploited to further enhance the system performance \cite{deploy}. Nevertheless, to reap the performance gain from the hybrid BS/terminal-integrated IS deployment,  an efficient CSI acquisition is needed, which is challenging to achieve due to more IS-associated channels involved. Moreover, under the hybrid deployment, the channel model becomes more sophisticated since the multi-hop reflection/transmission links exist, which increases the complexity and difficulty for IS passive reflection/transmission design.

\subsection{Integration of Active IS}

To mitigate the significant path loss due to the product-distance effects, a novel approach, called {\it active IS}, has emerged \cite{active_irs}, as depicted in Fig. \ref{future_picture}. This technology enables signal reflection and amplification simultaneously to achieve broader coverage, resulting in increased hardware costs and energy consumption compared to traditional passive IS architectures. Notice that the design in the current literature on active IS can be extended to the backside-IS architecture, while that for the frontside-IS architecture requires more research efforts, since the working mode of the active IS in the current literature is mainly reflective, while that in the frontside-IS architecture is transmissive. Furthermore, for the surrounding-IS architecture, the existence of double-reflection channels amplifies the noise introduced by active ISs, which complicates the CSI acquisition and IS reflection design. 

\subsection{Integration of Beyond Diagonal (BD)-IS}

In contrast to conventional ISs with diagonal phase-shift matrices, the BD-IS is equipped with scattering matrices that extend beyond diagonal configurations through inter-connections among IS elements \cite{beyond}. Owing to its flexible inter-element connections, BD-IS offers enhanced in manipulating waves intelligently and expanding coverage beyond what conventional IS architectures can achieve, as depicted in Fig. \ref{future_picture}. However, the deployment area at the BS is limited for additional circuits to realize inter-element connections in IS-integrated BS architectures, which poses a challenge for integrating BD-IS into BSs. In addition, the model of cascaded channels for BD-IS is different from that with conventional IS and only the  single-reflection channels have been studied in the current literature on BD-IS. Hence, effective modeling strategies for double-reflection channels should be further explored to facilitate channel estimation and IS reflection design in the surrounding-IS architecture with BD-IS.

\subsection{IS-Integrated BS for Integrated Sensing and Communication (ISAC)}

The IS-integrated BS for ISAC is a promising direction \cite{sensing}, since the extremely short distance between the BS's antenna array and IS can markedly decrease the product-distance path-loss of the cascaded channels for improving both the communication and sensing performances, as depicted in Fig. \ref{future_picture}. In this case, the design of IS passive reflective/transmissive coefficients must carefully balance communication and sensing performance. One viable approach is employing orthogonal time sharing to manage IS-assisted communication and sensing in separate time slots, optimizing their allocation. In addition,  exploring the utilization of double-reflection channels in the surrounding-IS architecture to enhance the trade-off between communication and sensing performance is crucial for future research.

\section{Conclusions}\label{conclusion}
In this article, we provided a comprehensive overview of the applications of IS-integrated BS in wireless communications. It was shown that the IS-integrated BS can significantly reduce the product-distance path-loss of the cascaded BS-IS-user channels, thus greatly improving the system performance. In addition, we presented three different IS-integrated BS architectures according to the integrated location of the IS, and discussed their design issues, including channel modelling, CSI acquisition, and IS passive reflection/transmission design. Furthermore, numerical results are provided to show that the surrounding-IS architecture achieves higher user rate performance than backside-IS and frontside-IS architectures under the same total surface aperture of the IS by exploiting the additional double-reflection channels, and the robust codebook-based design achieves better performance than other codebook-based designs owing to the smaller codebook size and slow adaptation of IS patterns. Finally, we pointed out open problems and promising directions for IS-BS/terminal integration in wireless networks to inspire future research. It is hoped that this article will serve as a useful and inspiring resource for future research on IS-integrated BS to unlock its full potential in 6G wireless communications.

\section*{Acknowledgment}
This work is supported in part by Advanced Research and Technology Innovation Centre (ARTIC) of National University of Singapore under Research Grant R-261-518-005-720, The Guangdong Provincial Key Laboratory of Big Data Computing, the National Natural Science Foundation of China (No. 62331022), and the Guangdong Major Project of  Basic and Applied Basic Research (No.  2023B0303000001).

\vspace{-70mm}

\begin{IEEEbiographynophoto}{Yuwei Huang} (yuweihuang@u.nus.edu) is a Ph. D. candidate with National University of Singapore, Singapore.
\end{IEEEbiographynophoto}
\vspace{-70mm}

\begin{IEEEbiographynophoto}{Lipeng Zhu}(zhulp@nus.edu.sg) is a Research Fellow with the Department of Electrical and Computer Engineering, National University of Singapore.
\end{IEEEbiographynophoto}
\vspace{-70mm}

\begin{IEEEbiographynophoto}{Rui Zhang}[F'17] (elezhang@nus.edu.sg) is a Professor with the School of Science and Engineering, The Chinese University of Hong Kong, Shenzhen, and Shenzhen Research Institute of Big Data. He is also with the ECE Department of National University of Singapore.
\end{IEEEbiographynophoto}

\end{document}